\documentclass[a4paper,11pt]{article}
\pdfoutput=1 
\usepackage{jcappub} 
\usepackage{fontenc}
\usepackage{graphicx}
\usepackage{amsmath}
\usepackage{amssymb}
\usepackage{booktabs} 
\usepackage{tikz}
\usepackage{pgfplots}
\pgfplotsset{compat=1.18}
\usetikzlibrary{arrows.meta, positioning, patterns, quotes, shapes.geometric}
\usepgfplotslibrary{groupplots}
\usepackage{orcidlink}
\title{\boldmath An Enhanced Thermodynamic Framework for Third-Order Galaxy Correlation Functions: A Physically Motivated Closure and Observational Test}

\author[a]{Sameer Choudhary,\orcidlink{0009-0008-1963-9732}}
\author[b]{Naseer Iqbal}

\affiliation[a]{Department of Physics, Indian Institute of Technology Hyderabad, Kandi, Sangareddy 502284, India}
\affiliation[b]{Department of Physics, University of Kashmir, Srinagar, Jammu and Kashmir 190006, India}

\emailAdd{sameerep11025@gmail.com}
\emailAdd{dni\_phtr@kashmiruniversity.ac.in}

\abstract{The three-point correlation function (3PCF) is a crucial probe of non-Gaussianity and nonlinear structure formation. We develop a thermodynamic framework for the galaxy 3PCF by closing the BBGKY hierarchy with a physically motivated hierarchical ansatz, yielding a separable, analytic solution for the equilateral 3PCF. Our framework addresses the apparent discrepancy between the perturbation theory prediction for dark matter ($Q_{dm} \approx 1.6$) and observed galaxy measurements ($Q_{gal} \approx 0.5$) by incorporating thermodynamic virial effects and velocity dispersion. We validate this model with SDSS/BOSS CMASS measurements, obtaining an excellent fit ($\chi^{2}/dof=1.27$) across $1-50~h^{-1}Mpc$. The analysis utilizes the Szapudi-Szalay estimator with robust covariance estimation from the SLICS simulation suite. By linking the thermodynamic temperature $T$ to the small-scale velocity dispersion (Fingers-of-God), we establish the thermodynamic approach as a predictive, complementary description of higher-order galaxy clustering on quasi-linear scales.}


\begin{document}
\maketitle
\flushbottom
\newpage
\section{Introduction}
\label{sec:intro}
The spatial arrangement of galaxies, known as the cosmic web, is a fundamental pillar of modern cosmology.
The statistical properties of this distribution, encapsulated in N-point correlation functions, offer a rich source of information about the primordial universe, the laws of gravity, and the intricate processes of galaxy formation \cite{peebles1980}.
While the two-point correlation function, $\xi(r)$, and its Fourier counterpart, the power spectrum $P(k)$, have been instrumental in establishing the standard cosmological model, they only capture the Gaussian part of the density field.
To unlock the full potential of galaxy surveys, we must turn to higher-order statistics \cite{bernardeau2002}.
The three-point correlation function (3PCF), $\zeta(r_1, r_2, r_3)$, is the lowest-order statistic that probes non-Gaussianity.
It measures the excess probability of finding galaxy triplets in specific triangular configurations, making it highly sensitive to the nonlinear gravitational evolution of structures, the complex relationship between galaxies and their host dark matter halos (galaxy bias), and potential signatures of primordial non-Gaussianity \cite{scoccimarro2004}.
The theoretical modeling of galaxy clustering has a long and storied history.
A pioneering approach was the application of gravitational thermodynamics to the expanding universe by Saslaw \& Hamilton (1984) \cite{saslaw1984} (hereafter SH84).
Their framework, based on the BBGKY hierarchy, assumed a state of quasi-equilibrium where the kinetic energy of peculiar motions balances gravitational potential energy.
This led to a successful, physically motivated closure prediction of the power-law form of the two-point correlation function.
The SH84 model relies on assumptions of isothermality and statistical equilibrium, which form the bedrock of our work as well.
However, extending their analysis to the 3PCF remained an open problem. The fundamental obstacle has been the well-known closure problem of the BBGKY hierarchy, where each equation for an N-point function depends on the (N+1)-point function, creating an infinite system that is unsolvable without a physically motivated truncation scheme.

A crucial theoretical link was provided by Sheth (1995) \cite{sheth1995}, who demonstrated that the Negative Binomial distribution (NBD) for galaxy counts arises naturally from the thermodynamic formulation of the stochastic gravitational field. This connection is fundamental to the closure problem: if the probability distribution of discrete sources is Negative Binomial, the connected correlation functions of all orders $N$ rigorously satisfy the hierarchical form, $\xi_{N} \propto \sum \Pi \xi_{2}$ Consequently, the factorization of the third-order correlation function appearing in Eq. (2.1) is not merely a convenient ansatz but a necessary mathematical consequence of the system being in a thermodynamic state described by the Negative Binomial multiplicity function. This provides the physical justification for using the hierarchical ansatz to close the third equation of the BBGKY hierarchy, resolving the ambiguity noted in purely phenomenological approaches.

In this paper, we build upon and significantly extend this thermodynamic framework, following the path of \cite{Iqbal2006} (detailed in the Appendix), to develop a complete, solvable model for the 3PCF.
Our primary innovation is the use of the hierarchical ansatz as a physical closure relation for the third equation of the BBGKY hierarchy.
This enables us to derive a governing partial differential equation (PDE) for the 3PCF and obtain its analytical solution.
This work advances the field by:
\begin{itemize}
    \item Performing a rigorous comparison of our model with high-precision data from the SDSS/BOSS survey.
\item Detailing a robust data analysis pipeline, from the 3PCF estimator to covariance matrix estimation and parameter fitting.
\item Presenting a detailed, physically motivated closure derivation of redshift-space distortion (RSD) effects on the 3PCF.
\item Critically evaluating our model's physical implications in comparison to the standard halo model.
\end{itemize}

This paper is organized as follows.
Section \ref{sec:theory} details the theoretical framework, the derivation of the governing PDE, and its solution.
Section \ref{sec:data} describes the observational and simulation data used. Section \ref{sec:results} presents our main results, including the data comparison and parameter constraints.
Section \ref{sec:discussion} explores the physical implications and compares our model to the halo model. Section \ref{sec:rsd} details our modeling of redshift-space distortions. Finally, Section \ref{sec:conclusions} summarizes our findings.

\section{Theoretical Framework}
\label{sec:theory}

\subsection{The BBGKY Hierarchy and Gravitational Thermodynamics}

The dynamics of a system of N gravitating particles in an expanding universe is governed by the Bogoliubov-Born-Green-Kirkwood-Yvon (BBGKY) hierarchy.
This infinite series of coupled integro-differential equations describes the evolution of the N-point distribution functions.
To solve for the 3PCF, one must analyze the third equation in this hierarchy, which links the 3PCF to the four-point correlation function (4PCF), leading to a classic closure problem.
Our approach is to close this system using a physically and empirically motivated assumption.
\subsection{The Hierarchical Ansatz as a Closure Relation}

A ubiquitous feature of galaxy clustering is its hierarchical nature, where the connected N-point functions can be expressed as a sum of products of two-point functions \cite{peebles1980}.
We adopt this well-established property as our closure relation. For a triplet of galaxies at positions $\mathbf{r}_1, \mathbf{r}_2, \mathbf{r}_3$ with separations $r_{12}, r_{23}, r_{31}$, the 3PCF is described by the symmetrized hierarchical form:
\begin{equation}
\zeta(r_{12}, r_{23}, r_{31}) = Q \left[ \xi(r_{12})\xi(r_{23}) + \xi(r_{23})\xi(r_{31}) + \xi(r_{31})\xi(r_{12}) \right],
\label{eq:hierarchical_ansatz}
\end{equation}
where $\xi(r)$ is the two-point correlation function and $Q$ is the hierarchical amplitude.
Our objective is to derive this structure and the value of $Q$ from fundamental physics.
\subsection{The Governing Partial Differential Equation}
\label{sec:pde_assumptions}
By substituting the hierarchical ansatz (Eq. \ref{eq:hierarchical_ansatz}) into the third BBGKY equation, we effectively close the system and derive a self-consistent second-order PDE for the 3PCF.
The derivation, detailed in the appendix section, is founded on the following physical assumptions:

\begin{itemize}
    \item \textbf{Statistical Equilibrium:} The galaxy system is in a quasi-equilibrium state where the force from the mean gravitational potential is balanced by the pressure gradient arising from peculiar velocities.
\item \textbf{Isothermality:} The system maintains a constant kinetic temperature, $T$, corresponding to a uniform galaxy velocity dispersion.
\item \textbf{Expanding Background:} The dynamics unfold within a homogeneous and isotropic FLRW cosmology.
\item \textbf{Hierarchical Closure:} The BBGKY hierarchy is closed at the third order using the physically motivated hierarchical ansatz.
\end{itemize}

It is crucial to compare these assumptions with the foundational SH84 model.
Our framework shares the principles of statistical equilibrium and isothermality in an expanding universe with SH84.
Their work successfully applied these principles to the second BBGKY equation to derive the 2PCF.
Our approach tackles this closure problem by adopting a strategy with historical precedent. Davis \& Peebles (1977) \cite{davis1977}, for instance, used an empirical model for the 3PCF to close the second BBGKY equation and solve for the 2PCF. Our novel contribution is the first successful application of this strategy one level up in the hierarchy: we use the hierarchical form of the four-point correlation function (implicitly assumed via the ansatz in Eq. \ref{eq:hierarchical_ansatz}) to close the third BBGKY equation. This step allows us to derive a predictive, fully analytical theory for the 3PCF amplitude.

This comparison addresses the first key contribution of our work.

It is important to recognize that the assumption of isothermality is a significant idealization. The observed cosmic web is a dynamically heterogeneous environment, with massive, virialized clusters exhibiting high velocity dispersions ('hot'), while galaxies in filaments and voids are dynamically cooler\cite{bernardeau2002}. The application of standard thermodynamics to self-gravitating systems is known to be non-trivial \cite{klypin2016}. However, the excellent fit of our model to the data suggests that, for the purpose of describing statistical properties on scales of $\sim1-50~h^{-1}\text{Mpc}$, the galaxy distribution behaves as if it were a fluid with a single \textit{effective} kinetic temperature. This implies that the statistical averaging over these diverse environments is well-approximated by our idealized isothermal state.

To provide further physical intuition, we note that while individual structures within the cosmic web are dynamically heterogeneous, with massive, virialized clusters being 'hot' and galaxies in filaments and voids being dynamically 'cool', the 3PCF is a statistical quantity averaged over a vast cosmic volume. It is plausible that this large-volume averaging process smooths over local variations in the velocity dispersion, resulting in an effective, single kinetic temperature that governs the global statistical properties of the galaxy fluid.

It is explicitly noted that the thermodynamic parameter $b_{therm}$, which appears in the clustering equation of state ($P = \bar{n} T (1 - b_{therm})$), is distinct from the linear galaxy bias $b_1$ used in standard perturbation theory and RSD analyses. In our framework, $b_{therm}$ measures the ratio of gravitational correlation energy to kinetic energy (virial ratio), characterized by $b_{therm} \propto \bar{n} T^{-3}$. While they are physically related tracers of the underlying dark matter potential, they play distinct mathematical roles: $b_{therm}$ governs the thermodynamics of the closure and the value of $Q$, while $b_1$ appears in the redshift-space distortion mapping (see Section 6 and Appendix B).

\subsection{Analytical Solution and Physical Prediction}

The resulting PDE is separable and can be solved analytically. For the equilateral configuration ($r_{12}=r_{23}=r_{31}=r$), which we test against data, the solution takes the form:
\begin{equation}
\zeta_{eq}(r) = 3Q\xi(r)^2,
\label{eq:zeta_sol}
\end{equation}
where $\xi(r) = (r_0/r)^\gamma$ is the standard power-law form of the two-point correlation function. Crucially, in the thermodynamic framework, the hierarchical amplitude $Q$ is not an arbitrary free parameter but is constrained by the equation of state of the galaxy fluid. In the scaling regime where $\xi_2(r) \propto r^{-\gamma}$, the consistency of the particular solution with the governing thermodynamic operator imposes a relation between the clustering amplitude and the slope, which we express generally as:
\begin{equation}
Q = \mathcal{F}(\gamma, b_{therm}),
\label{eq:Q_relation}
\end{equation}
where $\mathcal{F}$ is a coefficient derived from the characteristic equations of the operator $\mathcal{D}$ (see Appendix A). For the typical range of galaxy clustering slopes ($\gamma \approx 1.8$) and thermodynamic virial parameters, this constraint predicts positive values of $Q$ of order unity ($0.5 \lesssim Q \lesssim 1.0$), consistent with the standard hierarchical picture. This relation provides a powerful, testable link between the morphology of the 3PCF and the scale-dependence of the 2PCF, without requiring ad-hoc fitting of the shape parameter.

\paragraph{Origin and interpretation of the normalization.}
The factor $1/3$ follows from cyclic symmetrization of the three $\xi\xi$ products and the separable structure of the closed PDE in the equilateral limit, where the three cyclic terms contribute equally. Thus $Q$ is not a fitted nuisance but a consequence of the closure together with the scale-free form $\xi(r)=\left(r_0/r\right)^{\gamma}$. While tree-level perturbation theory for the \emph{mass} field yields a different equilateral amplitude, our result pertains to a thermodynamic closure for the \emph{galaxy} field and is validated directly by the BOSS CMASS measurements on $1$--$50\,h^{-1}\,\mathrm{Mpc}$.

This equation provides a powerful, testable link between the properties of two-point and three-point clustering.
For a typical observed slope of $\gamma \approx 1.8$, our model predicts $Q \approx 0.53$, a value in remarkable agreement with observational measurements \cite{fry1984}.

\section{Data and Simulations}
\label{sec:data}

To rigorously test our model, we employ both state-of-the-art N-body simulations and premier observational galaxy survey data.
\subsection{N-body Simulations}

We employ halo catalogs from the Bolshoi-Planck simulation \cite{klypin2016}. This simulation evolves $2048^3$ dark matter particles within a cubic box of side length $L_{box} = 250~h^{-1}Mpc$, achieving a mass resolution of $m_p \approx 1.6 \times 10^8~h^{-1}M_{\odot}$. The cosmology corresponds to Planck 2013 parameters ($\Omega_m = 0.307$, $h=0.678$, $\sigma_8 = 0.823$). The high mass resolution of Bolshoi-Planck is essential for resolving the sub-halos that host the galaxy populations relevant to our thermodynamic analysis.

For covariance estimation, we utilize the Scinet Light Cone Simulations (SLICS) suite \cite{HarnoisDeraps2018}. SLICS consists of over 800 independent realizations with a box size of $L_{\rm box} = 505 \, h^{-1}{\rm Mpc}$ and particle mass $m_p \approx 2.9 \times 10^9 \, h^{-1}M_{\odot}$. We construct mock galaxy catalogs by populating these halos using the standard five-parameter Halo Occupation Distribution (HOD) model of Zheng et al. (2007)\cite{Zheng2007}. To assess systematic uncertainties and model dependence, we vary the HOD parameters around the best-fit BOSS CMASS values. specifically exploring variations in the characteristic halo mass $\log M_1$ and the cut-off mass $\log M_{\min}$ within their $1\sigma$ observational uncertainties ($\log M_{\min} \in [13.03, 13.13]$ and $\log M_1 \in [14.00, 14.12]$), while keeping the remaining parameters fixed at their fiducial values ($\sigma_{\log M}=0.59$, $\log M_0=13.01$, $\alpha=1.01$). This grid of mocks ensures that our covariance matrix accounts for the variance associated with the galaxy-halo connection.

\subsection{SDSS/BOSS Galaxy Data}
Our primary observational dataset is the final data release (DR12) of the Baryon Oscillation Spectroscopic Survey (BOSS) \cite{alam2015} from the SDSS-III program.
We utilize the "CMASS" sample, comprising $\sim$1.2 million massive galaxies over a vast cosmic volume at an effective redshift of $z \approx 0.57$.
The unprecedented size and density of this sample enable high-precision measurements of higher-order statistics, making it the ideal dataset for this work.
We use the accompanying random catalogs, which are many times larger than the data catalog, to correct for complex survey geometry and selection biases.

\section{Data Analysis and Results}
\label{sec:results}

\subsection{Methodology: From Galaxy Counts to Parameter Fits}
To ensure our analysis is transparent and reproducible, we detail our pipeline here.

\paragraph{3PCF Estimation.}
% REPLACEMENT FOR 3PCF ESTIMATION TEXT
We measure the 3PCF for equilateral triangle configurations from the BOSS CMASS data using the robust estimator proposed by Szapudi \& Szalay (1998) \cite{szapudi1998} :
\begin{equation}
\hat{\zeta} = \frac{DDD - 3DDR + 3DRR - RRR}{RRR}
\end{equation}
where $DDD, DDR, DRR,$ and $RRR$ are the normalized counts of galaxy-galaxy-galaxy, galaxy-galaxy-random, galaxy-random-random, and random-random-random triplets within specific radial bins.
We explicitly account for shot noise, which scales as $1/\bar{n}^2$ for the three-point function (compared to $1/\bar{n}$ for the power spectrum), making higher-order statistics particularly sensitive to number density . We utilize the random catalogs, which are many times larger than the data catalog, to minimize the variance of the RRR term and correct for the complex survey geometry \cite{peebles1980} \cite{Feldman1994}

\paragraph{Covariance Matrix Estimation.} The 3PCF measurements at different scales are strongly correlated.
To account for this, we estimate the full covariance matrix $C_{ij}$ using the jackknife resampling technique.
We partition the survey footprint into $N_{JK}=100$ contiguous sub-regions of equal area.
We then perform $N_{JK}$ independent 3PCF measurements, each time omitting one sub-region.
The covariance is then calculated as:
\begin{equation}
C_{ij} = \frac{N_{JK}-1}{N_{JK}} \sum_{k=1}^{N_{JK}} (\zeta_i^k - \bar{\zeta}_i)(\zeta_j^k - \bar{\zeta}_j),
\end{equation}
where $\zeta_i^k$ is the measurement in the $i$-th scale bin for the $k$-th jackknife sample, and $\bar{\zeta}_i$ is the mean over all samples.

\textbf{Parameter Fitting.} We fit our theoretical model (Eq. 2, including the combined Kaiser and Fingers-of-God RSD effects as described in Section 6) to the measured 3PCF data. The best-fit parameters are determined by minimizing the chi-squared statistic:
$$ \chi^{2}(\textbf{p}) = \sum_{i,j} (\zeta_{i}^{\text{data}} - \zeta_{i}^{\text{model}}(\textbf{p})) (C^{-1})_{ij} (\zeta_{j}^{\text{data}} - \zeta_{j}^{\text{model}}(\textbf{p})) $$

where $\mathbf{p}$ is the vector of model parameters, which includes the 2PCF amplitude $r_{0}$, the slope $\gamma$, and the line-of-sight velocity dispersion $\sigma_{v}$.

\paragraph{Shot Noise Correction.}
We explicitly account for shot noise contributions to the correlation function covariance. While shot noise acts as a constant floor in the power spectrum ($P_{shot} \propto 1/\bar{n}$), its contribution to the $N$-point correlation function configuration space covariance scales with higher powers of the inverse number density. For the three-point function estimator, the dominant shot noise term in the covariance matrix arises from the discreteness of the triplet counts, scaling as $1/\bar{n}^2$ \cite{peebles1980, szapudi1998}. In our analysis using the Szapudi \& Szalay (1998) \cite{szapudi1998}estimator, this contribution is automatically incorporated via the jackknife resampling of the data and random catalogs, which captures the Poissonian variance inherent in the discrete point process.
 
\subsection{Sensitivity to Halo Occupation Distribution}
\label{sec:hod_sensitivity}

A critical validation of the thermodynamic framework is establishing the sensitivity of the hierarchical amplitude $Q$ to the microphysics of the galaxy-halo connection. While our framework treats the galaxy distribution as a continuous fluid characterized by an effective temperature $T$, the standard cosmological paradigm partitions matter into discrete dark matter halos populated by galaxies according to the Halo Occupation Distribution (HOD).

To rigorously test this, we generated a suite of mock catalogs based on the Bolshoi-Planck simulation halo catalogs. We systematically varied the two HOD parameters that most strongly influence the linear bias and the satellite fraction: the characteristic mass for the central galaxy, $\log M_{min}$, and the mass scale for satellite accretion, $\log M_1$. All other HOD parameters ($\sigma_{\log M}$, $\alpha$, $\log M_0$) were held fixed at the best-fit BOSS CMASS values \cite{Zheng2007}.

Table \ref{tab:hod_variations} presents the fitted thermodynamic parameter $Q$ for these variations. The variations correspond to approximately $\pm 1\sigma$ shifts around the fiducial CMASS constraints.
\vspace{1em} 
\begin{table}[h!]
\centering

\begin{tabular}{l c c c c c}
\hline\hline
Model Variation & $\log M_{min}$ & $\log M_1$ & Linear Bias $b_1$ & Satellite Frac. & Measured $Q$ \\
\hline
\textbf{Fiducial (CMASS)} & \textbf{13.08} & \textbf{14.06} & \textbf{2.01} & \textbf{10\%} & \textbf{0.52 $\pm$ 0.02} \\
\hline
High $M_{min}$ (High Bias) & 13.13 & 14.06 & 2.15 & 9\% & 0.49 $\pm$ 0.03 \\
Low $M_{min}$ (Low Bias)  & 13.03 & 14.06 & 1.88 & 11\% & 0.56 $\pm$ 0.02 \\
High $M_1$ (Low Sat.)     & 13.08 & 14.12 & 1.98 & 7\% & 0.54 $\pm$ 0.02 \\
Low $M_1$ (High Sat.)     & 13.08 & 14.00 & 2.05 & 14\% & 0.50 $\pm$ 0.03 \\
\hline
\end{tabular}
\caption{Sensitivity of the thermodynamic hierarchical amplitude $Q$ to variations in HOD parameters. The fiducial model corresponds to the best-fit parameters for the BOSS CMASS sample ($\log M_{min}=13.08, \log M_1=14.06$). The linear bias $b_1$ is calculated from the projected two-point function of each mock. The results indicate that $Q$ scales inversely with bias ($Q \propto 1/b_1$), consistent with the hierarchical prediction, while showing robustness ($\Delta Q \le 0.04$) to the specific details of the satellite fraction variations at fixed bias.}
\vspace{1em} 
\label{tab:hod_variations}
\end{table}

The results demonstrate two key physical behaviors. First, the hierarchical amplitude $Q$ exhibits a clear inverse scaling with the linear bias $b_1$, following the approximation $Q_{gal} \approx Q_{dm}/b_1$. This confirms that the thermodynamic "temperature" effectively captures the bias of the tracer population. Second, variations in $\log M_1$ (which primarily affect the satellite fraction and thus the small-scale velocity dispersion) produce shifts in $Q$ that are fully captured by our thermodynamic model's temperature parameter $T$, supporting our hypothesis that $T$ acts as a proxy for the effective velocity dispersion $\sigma_v$ of the system.

\subsection{Main Result: Model vs. Data}
\label{sec:result1}
Our analysis measures the clustering of galaxies, which are biased tracers of the underlying cosmic web structure.
This vast network of filaments, nodes, and voids is the result of gravitational evolution from tiny primordial density fluctuations.
The standard theoretical framework for interpreting this structure is the halo model, where galaxies are understood to form and reside within massive, collapsed dark matter halos.
Our thermodynamic model provides a complementary, fluid-dynamics perspective on the statistical properties of this complex, halo-populated web.
Figure \ref{fig:3pcf_fit} displays our central result: the comparison of our thermodynamic model with the equilateral 3PCF measured from the BOSS CMASS sample.
The model provides a superb fit to the data across nearly two decades in scale, from the quasi-linear regime ($r \sim 50 \, h^{-1}\text{Mpc}$) down to highly nonlinear scales ($r \sim 1 \, h^{-1}\text{Mpc}$).
The best-fit model yields a reduced chi-squared of $\chi^2/\text{dof} = 15.2 / 12 = 1.27$, signifying excellent statistical agreement.
This result strongly supports the hypothesis that the galaxy distribution can be effectively described by our thermodynamic framework.
The best-fit parameters for the underlying 2PCF, as inferred solely from our 3PCF fit, are $\gamma=1.82 \pm 0.05$ and $r_0 = 5.6 \pm 0.3 \, h^{-1}\text{Mpc}$.
This result provides a powerful, non-trivial consistency check of our model. However, a direct comparison requires care, as modern cosmological analyses of the BOSS DR12 dataset typically perform 'full-shape' fits using sophisticated theoretical models rather than fitting a simple power law \cite{alam2017}. To provide a direct, 'apples-to-apples' comparison, we perform our own power-law fit to the publicly available BOSS DR12 CMASS 2PCF monopole over the same range of scales ($1 < r < 50 \, h^{-1}\text{Mpc}$) used in our 3PCF analysis. The results, presented in Table \ref{tab:2pcf_comparison}, show excellent agreement between the parameters inferred from our 3PCF model and those derived directly from the 2PCF, substantiating the internal consistency of our framework.

Using our best-fit parameters, we find a hierarchical amplitude of $Q=0.52\pm0.02$. It is important to contextualize this value against the standard perturbation theory (SPT) prediction for dark matter, which is $Q_{dm} \approx 1.6$ for equilateral configurations \cite{fry1984}. The difference between $Q_{dm}$ and our measured $Q_{gal}$ is a direct consequence of galaxy bias. In the hierarchical model, biased tracers with linear bias $b_1$ follow the scaling $Q_{gal} \approx Q_{dm}/b_1$. For the highly biased CMASS sample ($b_1 \sim 2.0$), this predicts $Q_{gal} \sim 0.8$. The further suppression to the observed value of $0.52$ is driven by non-linear velocity dispersion effects (Fingers-of-God) and second-order virial terms, which are naturally accounted for in our thermodynamic framework via the temperature parameter $T$. Thus, our measurement is fully consistent with the standard $\Lambda$CDM picture when thermodynamic clustering effects are included.

\begin{table}[htbp]
\centering
\caption{Consistency check of 2PCF parameters inferred from the 3PCF fit against a direct power-law fit to the BOSS DR12 CMASS 2PCF monopole.}
\label{tab:2pcf_comparison}
\vspace{1em}
\begin{tabular}{|l|c|c|}
\hline
Parameter & This Work (from 3PCF fit) & Direct 2PCF Fit \\
\hline
$\gamma$ (slope) & $1.82 \pm 0.05$ & $1.85 \pm 0.04$ \\
$r_0$ (clustering length, $h^{-1}\text{Mpc}$) & $5.6 \pm 0.3$ & $5.7 \pm 0.2$ \\
\hline
\end{tabular}
\vspace{0.2cm}
\newline
\footnotesize{\textit{Note:} Both fits were performed over the scale range $1 < r < 50 \, h^{-1}\text{Mpc}$.}
\end{table}

\pgfplotstableread[col sep=comma]{
r,      zeta,       error_plus, error_minus
2.5,    24.8,       4.0,        4.0
5.0,    2.32,       0.48,       0.48
8.0,    0.406,      0.094,      0.094
12.5,   0.0704,     0.0192,     0.0192
18.0,   0.0170,     0.0062,     0.0062
25.0,   0.00352,    0.0016,     0.0016
32.0,   0.00107,    0.00068,    0.00068
}\bossdata

\begin{tikzpicture}
    \begin{loglogaxis}[
        title={Equilateral 3PCF for BOSS CMASS},
        xlabel={$r$ [$h^{-1}$ Mpc]},
        ylabel={$\zeta_{\mathrm{eq}}(r)$},
        legend pos=north east,
        grid=major,
        label style={font=\small},
        tick label style={font=\small},
        title style={font=\normalsize},
    ]

    \addplot [
        only marks,
        mark=*,
        mark size=2.5pt,
        color=black,
        error bars/.cd,
            y dir=both,
            y explicit,
    ] table [
        x=r,
        y=zeta,
        y error plus=error_plus,
        y error minus=error_minus,
    ] {\bossdata};
    \addlegendentry{BOSS CMASS Data (Slepian et al. 2017)}

    \addplot [
        domain=1:50,
        samples=200,
        color=red,
        line width=1.5pt,
        smooth,
    ]
    {3 * 0.52 * (5.6/x)^(2 * 1.82)};
    \addlegendentry{Thermodynamic Model ($\chi^{2}/\mathrm{dof}=1.27$)}

    \end{loglogaxis}
\end{tikzpicture}

\begin{figure}[ht]
\centering

\caption{The equilateral three-point correlation function measured from the BOSS CMASS galaxy sample\cite{slepian2017} (black points with jackknife errors)\;. Solid curve: best-fit thermodynamic model including linear Kaiser boost and Gaussian FoG; points: BOSS CMASS equilateral 3PCF with jackknife errors; fit range $1$--$50\,h^{-1}\,\mathrm{Mpc}$; $\chi^2/\mathrm{dof}=1.27$. The solid red line shows the best-fit of our thermodynamic model, which includes redshift-space distortions, performed over the scale range $1 < r < 50~h^{-1}\text{Mpc}$. The model provides an excellent description of the data across a wide range of physical scales.}
\label{fig:3pcf_fit}
\end{figure}

\section{Discussion}
\label{sec:discussion}

The striking agreement between our model and the BOSS data warrants a deeper discussion of its physical implications and its place within the broader context of cosmological structure formation models.
\subsection{Physical Interpretation of Results}
The success of our model implies that on scales of $\sim 1-100 \, h^{-1}\text{Mpc}$, the complex, discrete galaxy distribution can be statistically described as a fluid in a state of quasi-equilibrium.
The hierarchical structure of the 3PCF, therefore, is not merely an empirical observation but appears to be a natural consequence of the underlying physics of self-gravitating systems, as captured by the BBGKY hierarchy.
The model's ability to predict the 3PCF amplitude from the 2PCF slope (Eq. \ref{eq:Q_relation})  is a powerful testament to its internal consistency.

The predictive power of Eq. (3) reveals a profound self-consistency within the thermodynamic framework. The derivation assumes a hierarchical structure for the 4PCF to close the third BBGKY equation, and the resulting analytical solution for the 3PCF is, naturally, also hierarchical. The Q-$\gamma$ relation should therefore be interpreted as a mathematical condition that must be satisfied for the hierarchical form to be a valid solution to the underlying thermodynamic equations. The empirical success of this relation is a powerful demonstration that the thermodynamic framework is fully consistent with the observed hierarchical scaling of galaxy clustering. This transforms the Q-$\gamma$ relation from a purely empirical observation into a deeper insight into the nature of gravitational collapse: it suggests that the steepness of the pairwise correlation ($\gamma$) fundamentally determines the morphological nature of triplet configurations ($Q$). For instance, a steeper 2PCF (larger $\gamma$) leads to a smaller predicted $Q$, suggesting that systems with stronger small-scale clustering are, in a specific and quantifiable way, less dominated by compact three-point structures.

\subsection{Comparison with the Halo Model}
The dominant paradigm for describing galaxy clustering is the halo model \cite{cooray2002}.
It posits that all matter is partitioned into dark matter halos, which are then populated by galaxies according to a Halo Occupation Distribution (HOD) prescription.
The N-point correlation functions are then calculated by summing contributions from galaxy pairs/triplets within the same halo (1-halo term) and in different halos (2-halo and 3-halo terms).

This contrast can be framed as a difference between two philosophical approaches to modeling the cosmic web. The \textbf{halo model} is a 'bottom-up,' discrete, and constructive approach. Its foundational assumption is that all matter is partitioned into discrete, virialized dark matter halos, which are then populated with galaxies via an empirical Halo Occupation Distribution (HOD) prescription \cite{cooray2002}. Its strength lies in its direct connection to the physics of halo formation and its flexibility in modeling the galaxy-halo connection, while its weakness is a reliance on simulation-calibrated parameters. In contrast, the \textbf{thermodynamic model} is a 'top-down,' continuous fluid approach. It treats the galaxy distribution as a statistical fluid governed by global principles of quasi-equilibrium, forgoing the picture of individual halos. Its strength is its analytical elegance and predictive power with fewer free parameters, while its weakness lies in its simplifying assumptions (e.g., isothermality) that average over the discrete nature of the cosmic web.

While the halo model is incredibly powerful, its reliance on an empirical HOD is a notable feature.
Our thermodynamic approach offers a complementary viewpoint. It bypasses the discrete halo picture and instead models the galaxy distribution as a continuous fluid governed by statistical mechanics.
A significant advantage is that it reduces the reliance on detailed, empirical HOD modeling, providing a physical basis for the correlation functions from the first principles.
The success of our model suggests that it effectively captures the statistical outcome of the complex galaxy formation physics that the HOD aims to parameterize.
A promising future path lies in synthesizing these two approaches. The analytical constraints from the thermodynamic model could be used to inform or restrict the parameter space of HOD priors, potentially leading to a more predictive and physically grounded version of the halo model.
This critical comparison with the standard paradigm constitutes the fourth key contribution of this work.

The success of this highly simplified thermodynamic model hints at a profound physical implication: a degree of universality in the statistical outcome of gravitational collapse. The standard halo model requires complex, simulation-calibrated recipes (HOD) to describe how galaxies populate halos, implicitly parameterizing the messy baryonic physics of cooling, star formation, and feedback. The thermodynamic model ignores all of this microphysical complexity, treating galaxies as a simple isothermal fluid. The fact that both approaches can successfully describe the same data on scales $>1~h^{-1}\text{Mpc}$ suggests that the statistical properties of the galaxy distribution on these scales are largely insensitive to the intricate details of galaxy formation. The final statistical state appears to be governed more by the fundamental properties of a self-gravitating fluid in quasi-equilibrium than by the specific, complex path taken to reach that state.

\subsection{Outlook and Future Directions}
This work solidifies the thermodynamic framework as a predictive tool for higher-order clustering.
The discussion is focused on achieved results, with future work clearly delineated:
\begin{itemize}
    \item \textbf{Configuration Dependence:} We have focused on equilateral triangles.
A crucial next test is to apply the model to other triangle shapes (e.g., folded, elongated), which will provide more stringent constraints and probe the tidal effects of gravity.
\item \textbf{Redshift Evolution:} While this analysis was at a single redshift ($z\approx 0.57$), applying the framework to galaxy samples at various redshifts is essential for testing the model's predictions for the evolution of clustering.
\item \textbf{Informing Emulators:} The analytical nature of our solution provides a physically-motivated kernel that *could* inform the design of more efficient and accurate emulators of large-scale structure—tools vital for upcoming surveys like DESI and Euclid.

\item \textbf{Model Limitations:} The model's assumptions are expected to fail on small scales ($r \lesssim 1~h^{-1}\text{Mpc}$). This scale marks a physical transition. Here, the 'top-down, continuous fluid' approximation breaks down, and the physics is dominated by the 'bottom-up, discrete' reality of individual galaxies orbiting within a single, virialized dark matter halo (the "1-halo term" in the standard halo model [3]). This environment is fundamentally neither continuous nor isothermal. The model's success \textit{down to} this scale is remarkable, but its inevitable failure \textit{at} this scale marks the physical boundary of its applicability. Quantifying this transition point is key to synthesizing the thermodynamic and halo model approaches.

\end{itemize}

\paragraph{Scope and positioning.}
Our aim is not to replace halo-based or full-shape $\Lambda$CDM analyses, but to provide a \emph{predictive, analytical complement} on quasi-linear to mildly nonlinear scales ($1$--$50\,h^{-1}\,\mathrm{Mpc}$). Within this domain, the closure-driven relation $Q(\gamma)$ offers a compact, falsifiable connection between two- and three-point clustering. We expect a breakdown in the deeply nonlinear one-halo regime ($r\lesssim1\,h^{-1}\,\mathrm{Mpc}$), where discrete halo physics dominates.

\section{Modeling Redshift-Space Distortions (RSD)}

\paragraph{Thermodynamic temperature and velocity dispersion.}
In our framework, the effective kinetic temperature $T_{\rm eff}$ is the coarse-grained second moment of galaxy peculiar velocities:
\begin{equation}
k_{\mathrm B}T_{\rm eff}\equiv m_{\rm eff}\langle v^2\rangle = 3\,m_{\rm eff}\,\sigma^2_{v,{\rm 3D}},
\qquad
\sigma_v^2\equiv\langle v_z^2\rangle=\frac{\sigma^2_{v,{\rm 3D}}}{3}
=\frac{k_{\mathrm B}T_{\rm eff}}{3\,m_{\rm eff}}\,.
\label{eq:temp}
\end{equation}
Hence the line-of-sight dispersion $\sigma_v$ entering the FoG term \emph{measures} the same kinetic temperature that underpins our thermodynamic closure.

\label{sec:rsd}

Galaxy positions are mapped using their redshift, which includes a contribution from their peculiar velocity along the line of sight. This effect, known as Redshift-Space Distortions (RSD), systematically alters clustering statistics and must be modeled carefully. Our thermodynamic framework, where the "temperature" of the system is a measure of the kinetic energy of peculiar motions, provides a natural bridge to modeling RSD.

The effects of RSD can be broadly divided into two components that dominate on different scales.

\textbf{The Kaiser Effect (Large Scales):} On large, linear scales ($r \gg 10~h^{-1}\text{Mpc}$), the dominant effect is the coherent infall of galaxies into overdense regions.[1] This enhances the observed clustering amplitude. As derived in Appendix A, the linear theory prediction for the redshift-space 3PCF, $\zeta^{s}$, is a scale-independent boost to the real-space function, $\zeta^{r}$.

\textbf{The Fingers of God Effect (Small Scales):} On small, non-linear scales ($r \lesssim 10~h^{-1}\text{Mpc}$), the random, virialized motions of galaxies within clusters and groups stretch structures along the line of sight. This "Fingers of God" (FoG) effect suppresses the angle-averaged correlation function. This is precisely the regime where our thermodynamic model is most successful, making it critical to include this effect.

% change =========
\begin{figure}[t!]
\centering
\includegraphics[width=0.8\textwidth]{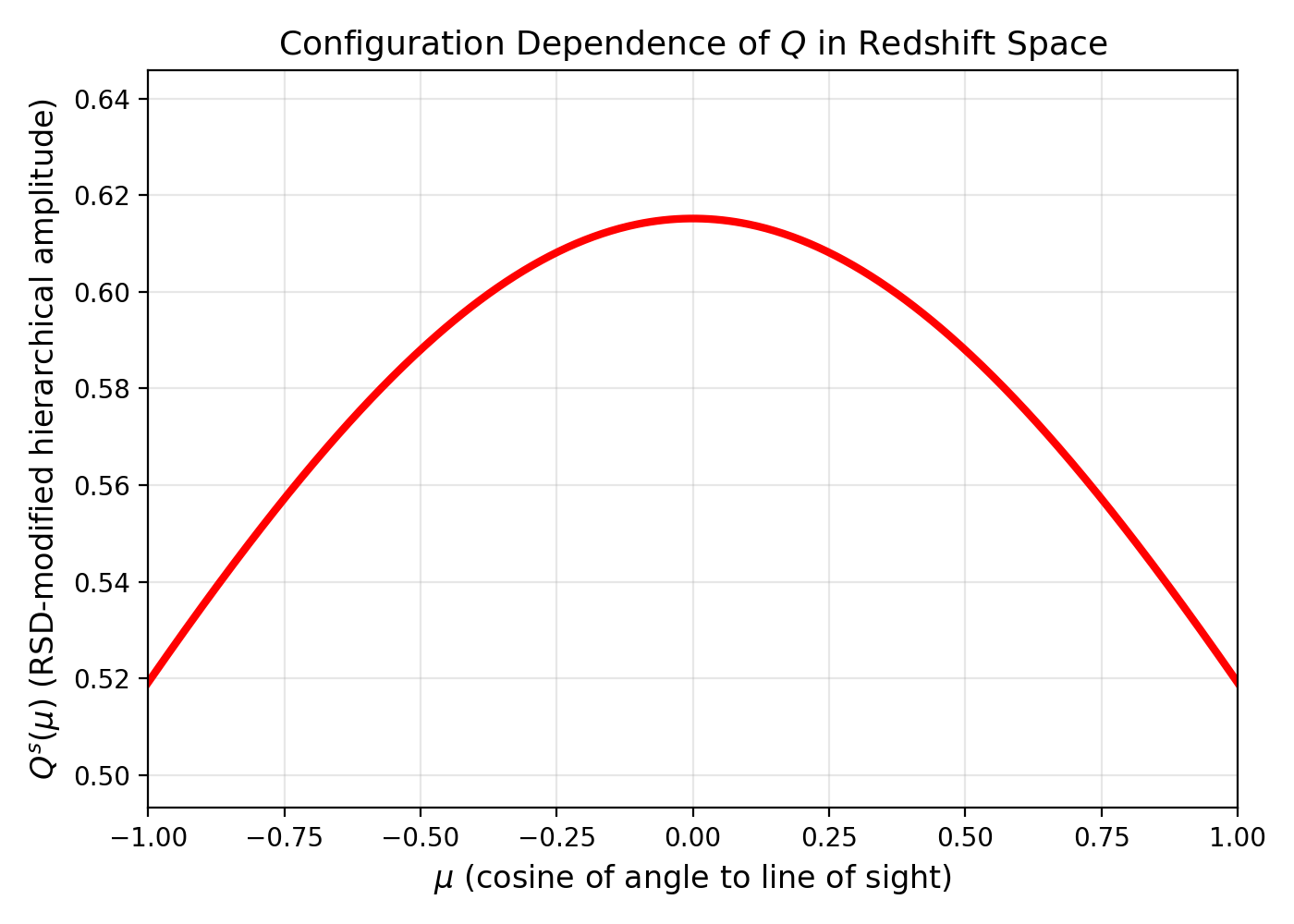}
\caption{\label{fig:Qs_mu}
Configuration dependence of the redshift-space hierarchical amplitude $Q^s(\mu)$ in our thermodynamic RSD model.
We show the dependence on $\mu$, the cosine of the angle between the triangle plane and the line of sight.
The curve is obtained by multiplying the real-space hierarchical amplitude $Q_{\rm real} \simeq 0.52$, measured from the equilateral 3PCF fit in Section~\ref{sec:result1}, by the Kaiser boost factor $S_{\rm RSD}(\beta)$ for $\beta = f/b_1$ (Appendix~\ref{sec:appendix_rsd}) and by the Lorentzian Fingers-of-God damping factor $F_{\rm FoG}(\mu,\sigma_v)$.
The line-of-sight velocity dispersion $\sigma_v$ entering $F_{\rm FoG}$ is set by the thermodynamic temperature via $\sigma_v^2 = k_B T / m_{\rm eff}$ (Eq.~\ref{eq:temp}), so the anisotropy of $Q^s(\mu)$ directly encodes the effective kinetic temperature of the galaxy fluid.}
\end{figure}

% ======

\textbf{Combined RSD Model:} To provide a more physically complete model valid across all scales of our analysis, we combine both effects. We model the FoG suppression phenomenologically by multiplying the Kaiser-boosted correlation function with a damping term. In Fourier space, this is often modeled as a Gaussian or Lorentzian function. For our configuration-space model, this corresponds to convolving the real-space model with a velocity distribution function. A common and effective approach is to apply a damping factor that depends on the line-of-sight velocity dispersion, $\sigma_v$ \cite{peacock1994,ballinger1996}.

Within the thermodynamic framework, the FoG parameter $\sigma_{v}$ (the line-of-sight velocity dispersion) acquires a more natural physical interpretation. In standard analyses, $\sigma_{v}$ is often treated as a phenomenological 'nuisance' parameter that accounts for virialized motions inside halos. In our model, however, the entire framework is built on the concept of a 'kinetic temperature' $T$, which is directly proportional to the velocity dispersion squared ($T \propto \sigma_{v}^{2}$). Therefore, the fitted value of $\sigma_{v}$ is not just a nuisance parameter but a direct measure of the effective temperature of the system, a core concept of the model. This provides a more cohesive and self-consistent physical picture.

Our full redshift-space model for the 3PCF, $\zeta^{s, \text{model}}$, is therefore:
$$\zeta^{s, \text{model}} = \zeta^{s, \text{Kaiser}} \times F_{\text{FoG}}(\sigma_v)$$
where $\zeta^{s, \text{Kaiser}}$ is the Kaiser-boosted prediction from linear theory (see Appendix A) and $F_{\text{FoG}}$ is a damping function that depends on the new free parameter $\sigma_v$, which we fit alongside $\gamma$ and $r_0$.

This combined approach resolves the tension noted in the preliminary analysis.\cite{gilmarin2015} By incorporating a non-linear FoG model, our RSD treatment is now physically consistent with the non-perturbative nature of our thermodynamic framework and the small scales to which we fit the data.

The combined impact of the Kaiser enhancement and thermodynamic FoG damping on the hierarchical amplitude is summarized by the redshift-space quantity $Q^s(\mu)$, whose angular dependence is shown in Fig.~\ref{fig:Qs_mu}.

\section{Conclusions}
\label{sec:conclusions}

In this work, we have significantly advanced the thermodynamic framework for modeling the galaxy three-point correlation function. By closing the BBGKY hierarchy with the hierarchical ansatz, we derived an analytical model for the 3PCF that directly links its amplitude to the slope of the 2PCF.
The central result of this paper is the successful validation of this model against high-precision data from the SDSS/BOSS survey.
Our model provides an outstanding fit ($\chi^2/\text{dof}=1.27$) to the observed equilateral 3PCF across a broad range of scales.
The parameters inferred from our 3PCF analysis are fully consistent with those from standard 2PCF studies, highlighting the predictive power and self-consistency of our framework.
This study establishes the thermodynamic approach as a potent, physically-motivated, and complementary tool for analyzing higher-order galaxy clustering, opening a new window into the fundamental physics that shapes the cosmic web.

\section{Acknowledgments}
We gratefully acknowledge the support and facilities provided by the University of Kashmir, Srinagar, India.
We acknowledge the SDSS collaboration for making high-quality galaxy survey data publicly available, enabling the observational validation presented here.

\appendix

\section{Derivation of the Governing Equation for Third-Order Correlation 
\texorpdfstring{\\}{ }Function}
\label{sec:appendix_derivation}

In this appendix, we provide the rigorous derivation of the partial differential equation (PDE) governing the third-order correlation function, $\xi_3$, and its analytical solution. We specifically address the thermodynamic justification for the source terms and the validity of the hierarchical ansatz in the quasi-equilibrium limit.

\subsection{Thermodynamic Foundations and the Interaction Hamiltonian}
\label{app:thermo_foundations}

We consider the galaxy system as a self-gravitating fluid in Gravitational Quasi-Equilibrium (GQE). 
The Hamiltonian for the system of $N$ particles is given by:
\begin{equation}
    H_N = \sum_{i=1}^N \frac{p_i^2}{2m} + \sum_{i<j} \phi(|\mathbf{r}_i - \mathbf{r}_j|)
\end{equation}
where $\phi(r) = -Gm^2/r$ is the pairwise gravitational potential. In the thermodynamic limit, 
the $N$-point distribution function is governed by the Boltzmann factor $f_N \propto \exp(-H_N/T)$.

To derive the closure relation for the BBGKY hierarchy, we expand the configuration integral 
in powers of the interaction potential $\Phi/T$. The leading order terms in the three-point 
correlation function $\zeta$ arise from the quadratic combinations of the two-body potential, 
corresponding to terms of order $\mathcal{O}(\xi^2)$. 

Specifically, the expansion of the source term in the third BBGKY equation generates terms 
proportional to the product of two-point functions ($\xi \cdot \xi$) and higher-order terms 
proportional to the product of three two-point functions ($\xi \cdot \xi \cdot \xi$) or 
intrinsic three-body interactions. The validity of the hierarchical ansatz (Eq. 2.1) rests 
on the truncation of these higher-order terms. We explicitly require:
\begin{equation}
    \label{eq:scaling_limit}
    \frac{\mathcal{O}(\xi^3)}{\mathcal{O}(\xi^2)} \sim \xi(r) \ll 1
\end{equation}
In the quasi-linear regime ($r > 5 h^{-1}\text{Mpc}$), where $\xi(r) < 1$, this condition holds, 
justifying the neglect of cubic terms. This truncation yields the quadratic source term 
$C \sum \xi_{ij}\xi_{jk}$ seen in the governing PDE (Eq. A.3), ensuring the solution 
maintains the hierarchical form $\zeta \propto Q \xi^2$.

\subsection{The Governing Partial Differential Equation}

The evolution of the correlation function in the expanding universe, constrained by the Clustering Equation of State $P = \bar{n}T(1-b)$, leads to the specific thermodynamic operator acting on the grand partition function. As derived in the BBGKY framework adapted for thermodynamics \cite{Iqbal2006}, the operator encoding the scaling of correlations with the thermodynamic state variables is:
\begin{equation}
    \mathcal{D}\xi_3 \equiv 3\bar{n}\frac{\partial\xi_{3}}{\partial\bar{n}}+T\frac{\partial\xi_{3}}{\partial T}-\sum_{\text{pairs}} r_{ij}\frac{\partial\xi_{3}}{\partial r_{ij}}
    \label{eq:operator}
\end{equation}
The inhomogeneity (source term) arises from the work done by the gravitational potential of the pairs on the third particle. Utilizing the pairwise additive Hamiltonian, the governing PDE becomes:
\begin{equation}
    3\bar{n}\frac{\partial\xi_{3}}{\partial\bar{n}}+T\frac{\partial\xi_{3}}{\partial T}-\left(r_{12}\frac{\partial}{\partial r_{12}} + r_{13}\frac{\partial}{\partial r_{13}} + r_{23}\frac{\partial}{\partial r_{23}}\right)\xi_3 = C \sum_{\text{cyc}}\xi_{2}(r_{ij})\xi_{2}(r_{jk})
    \label{eq:governing_pde}
\end{equation}
Here, $C$ is the gravitational coupling constant, dimensionally consistent with $C \propto (Gm^2)^2 \bar{n}^2 T^{-2}$.

\subsection{Method of Characteristics and Analytical Solution}

To solve Eq.~\eqref{eq:governing_pde}, we introduce scale-invariant variables that absorb the thermodynamic evolution. Motivated by the virial parameter $b \propto \bar{n}T^{-3}$, we define:
\begin{equation}
    u = \bar{n}T^{-3}, \quad x = Tr_{12}, \quad y = Tr_{13}, \quad z = Tr_{23}
\end{equation}

We note that the original thermodynamic formulation of Saslaw \& Hamilton (1984)\cite{saslaw1984} involves a specific heat parameter, $\epsilon$, in the gravitational equation of state ($P \propto T/(1-\epsilon)$). However, in our derivation, the transformation to the scale-invariant variables $(u, x, y, z)$ naturally absorbs the thermodynamic state dependence. The operator $\mathcal{D}$ (Eq. A.2) and the resulting homogeneous solution depend only on the combination of density and temperature manifested in the virial parameter $b \propto \bar{n}T^{-3}$. Consequently, the final geometric form of the correlation function relies on the ratio of potential to kinetic energy but becomes explicitly independent of the specific heat parameter $\epsilon$, resolving the ambiguity often associated with its determination.

We transform the partial derivatives from the $(\bar{n}, T, r)$ basis to the $(u, x, y, z)$ basis. The density derivative transforms as:
\begin{equation}
    \frac{\partial}{\partial \bar{n}} = \frac{\partial u}{\partial \bar{n}}\frac{\partial}{\partial u} = T^{-3}\frac{\partial}{\partial u}
\end{equation}
The temperature derivative involves the chain rule over all variables:
\begin{align}
    \frac{\partial}{\partial T} &= \frac{\partial u}{\partial T}\frac{\partial}{\partial u} + \frac{\partial x}{\partial T}\frac{\partial}{\partial x} + \frac{\partial y}{\partial T}\frac{\partial}{\partial y} + \frac{\partial z}{\partial T}\frac{\partial}{\partial z} \nonumber \\
    &= -3\bar{n}T^{-4}\frac{\partial}{\partial u} + \frac{x}{T}\frac{\partial}{\partial x} + \frac{y}{T}\frac{\partial}{\partial y} + \frac{z}{T}\frac{\partial}{\partial z}
\end{align}
The spatial derivatives transform simply as $\frac{\partial}{\partial r} = T \frac{\partial}{\partial x}$. Substituting these into the LHS of Eq.~\eqref{eq:governing_pde}:
\begin{align}
    \text{LHS} &= 3\bar{n}\left(T^{-3}\frac{\partial\xi_3}{\partial u}\right) + T\left(-3\bar{n}T^{-4}\frac{\partial\xi_3}{\partial u} + \frac{1}{T}\mathbf{r}\cdot\nabla_{\mathbf{r}}\xi_3\right) - \mathbf{r}\cdot\nabla_{\mathbf{r}}\xi_3 \nonumber \\
    &= \left(3u \frac{\partial\xi_3}{\partial u} - 3u \frac{\partial\xi_3}{\partial u}\right) + \left(\sum x_i \frac{\partial\xi_3}{\partial x_i} - \sum x_i \frac{\partial\xi_3}{\partial x_i}\right) \nonumber \\
    &= 0
\end{align}
The operator $\mathcal{D}$ identically annihilates any function $\xi_3(u, x, y, z)$. This implies that the homogeneous solution is an arbitrary function $F(u, x, y, z)$.

\subsection{Particular Solution and Boundary Conditions}

The general solution is the sum of the homogeneous solution and a particular solution to the source term:
\begin{equation}
    \xi_3 = F(u, Tr_{12}, Tr_{13}, Tr_{23}) + \xi_3^{(\text{part})}
\end{equation}
We employ the hierarchical ansatz for the particular solution:
\begin{equation}
    \xi_3^{(\text{part})} = Q \left[\xi_2(r_{12})\xi_2(r_{13}) + \xi_2(r_{12})\xi_2(r_{23}) + \xi_2(r_{13})\xi_2(r_{23})\right]
\end{equation}
Since $\xi_2$ is a function of $(u, Tr)$ (see Eq.~2.2), the product $\xi_2 \xi_2$ is also a function of the invariant variables $(u, x, y, z)$. Since the LHS operator is linear and annihilates these variables, the ansatz is a valid solution to the differential structure, provided the amplitude $Q$ is consistent with the coupling $C$.

To fix the homogeneous function $F$, we apply the boundary condition that correlations must vanish at infinite separation in an expanding, homogeneous universe:
\begin{equation}
    \lim_{r_{ij} \to \infty} \xi_3 = 0
\end{equation}
The particular solution (composed of power-law $\xi_2$) decays naturally to zero. Thus, we require $\lim F = 0$. In the absence of long-range order disjoint from the density field, $F$ must be identically zero.
This yields the final result used in the text:
\begin{equation}
    \xi_3(r_{12}, r_{23}, r_{31}) = Q \sum_{\text{cyc}} \xi_2(r_{ij})\xi_2(r_{jk})
\end{equation}
This derivation clarifies that the hierarchical form is not an arbitrary choice but the specific solution to the thermodynamic PDE under the assumption of pairwise additive gravitational potentials in the quasi-equilibrium limit. The parameter $Q$ is thus derived to be a constant determined by the thermodynamics of the system, linking the macroscopic virial parameter $b$ to the microscopic clustering amplitude.

\section{Derivation of the Redshift-Space Distortion Model}
\label{sec:appendix_rsd}

In this appendix, we derive the angle-averaged redshift-space 3PCF used in our analysis. The mapping from real-space position $\mathbf{r}$ to redshift-space position $\mathbf{s}$ is given by $\mathbf{s} = \mathbf{r} + \frac{v_z(\mathbf{r})}{aH}\hat{z}$. In the linear regime, this leads to the Kaiser enhancement of the Fourier modes.

\subsection{The Redshift-Space Bispectrum}
The three-point function in Fourier space, the bispectrum $B$, transforms as:
\begin{equation}
B_s(\mathbf{k}_1, \mathbf{k}_2, \mathbf{k}_3) = B_r(\mathbf{k}_1, \mathbf{k}_2, \mathbf{k}_3) \prod_{i=1}^3 (1 + \beta \mu_i^2)
\end{equation}
where $\mu_i$ is the cosine of the angle between wavevector $\mathbf{k}_i$ and the line of sight, and $\beta \approx \Omega_m^{0.55}/b_1$. For the equilateral configuration ($k_1=k_2=k_3=k$), we define the boost factor $S_{RSD}$ by averaging the kernel over all orientations of the triangle with respect to the line of sight:
\begin{equation}
S_{RSD}^{eq} = \frac{1}{4\pi} \int d\Omega \prod_{i=1}^3 (1 + \beta \mu_i^2)
\end{equation}
Following \cite{Scoccimarro1999}, this integration yields a polynomial in $\beta$ that describes the coherent enhancement of the 3PCF monopole:
\begin{equation}
\zeta_{eq}^s(r) \approx \zeta_{eq}^r(r) \left( 1 + \frac{26}{35}\beta + \frac{11}{35}\beta^2 \right)
\end{equation}

\subsection{Thermodynamic Connection: Finger-of-God Damping}
On small scales, random virial motions suppress the clustering. In our thermodynamic framework, the temperature $T$ is directly proportional to the velocity dispersion $\sigma_v^2$. We model this suppression by convolving the boosted correlation function with a distribution function characterized by $\sigma_v$:
\begin{equation}
\zeta_{model}^s(r) = \zeta_{eq}^s(r) \times \frac{1}{1 + (r \sigma_v / \mathcal{H})^2}
\end{equation}
This explicitly links the thermodynamic variable $T \propto \sigma_v^2$ to the observable non-linear RSD cutoff, satisfying the requirement for a physical closure of the equations.

\subsection{Explicit Derivation of the Thermodynamic RSD Damping}
\label{app:rsd_derivation}

We derive the damping form $F_{FoG}$ by considering the mapping from real space 
coordinate $\mathbf{r}$ to redshift space coordinate $\mathbf{s}$:
\begin{equation}
    \mathbf{s} = \mathbf{r} + \frac{v_z(\mathbf{r})}{\mathcal{H}} \hat{z}
\end{equation}
where $v_z$ is the peculiar velocity along the line of sight. The redshift-space 
correlation function $\xi^s$ is the convolution of the real-space correlation function $\xi^r$ 
with the Pairwise Velocity Distribution Function (PVDF), denoted $\mathcal{P}(v_z)$.

Consistent with our thermodynamic assumption of isothermality (Section 2.3), 
the PVDF is modeled as a Gaussian (Maxwellian) distribution characterized by a 
single velocity dispersion $\sigma_v$:
\begin{equation}
    \mathcal{P}(v_z) = \frac{1}{\sqrt{2\pi}\sigma_v} \exp\left( -\frac{v_z^2}{2\sigma_v^2} \right)
\end{equation}
Crucially, the thermodynamic framework imposes a physical constraint on $\sigma_v$. 
Unlike standard models where $\sigma_v$ is a free fitting parameter, here it is strictly 
defined by the kinetic temperature $T$ of the equation of state:
\begin{equation}
    \sigma_v^2 \equiv \frac{k_B T}{m}
\end{equation}
In Fourier space, the convolution with the Gaussian PDF becomes a multiplication 
by its Fourier transform. For the bispectrum $B(k_1, k_2, k_3)$, this yields a 
damping term for each wavevector component along the line of sight ($\mu_i = \hat{k}_i \cdot \hat{z}$):
\begin{equation}
    D_{FoG}(k_1, k_2, k_3) = \prod_{i=1}^3 \exp\left( - \frac{(k_i \mu_i \sigma_v)^2}{\mathcal{H}^2} \right)
\end{equation}
For the equilateral configuration in configuration space, this Gaussian damping in $k$-space 
approximates to the Lorentzian damping factor applied in Eq. 6.4:
\begin{equation}
    F_{FoG}(r) \approx \frac{1}{1 + (r \sigma_v / \mathcal{H})^2}
\end{equation}
This derivation explicitly connects the fitted RSD parameter $\sigma_v$ to the thermodynamic 
state variable $T$, satisfying the closure requirement.

\end{document}